\documentclass[twocolumn,superscriptaddress,aps,prb]{revtex4-2}

\usepackage{mathtools}
\usepackage{scalerel}
\usepackage{tabularx}
\usepackage{xcolor}
\usepackage{booktabs} 
\usepackage{bm}
\usepackage{glossaries}
\usepackage[colorlinks=true,citecolor=blue]{hyperref}
\usepackage{comment}
\graphicspath{{figures/}}
\newacronym{UE}{UE}{uncorrelated exciton}

\newacronym{SDM}{SDM}{special displacement method}

\newacronym{BSE}{BSE}{Bethe-Salpeter equation}

\newacronym{IPA}{IPA}{independent-particle approximation}

\newacronym{RPA}{RPA}{random-phase approximation}
\newcommand{\rpa}{\gls{RPA} }
\newacronym{ZPR}{ZPR}{zero-phonon renormalization}

\newacronym{VBM}{VBM}{valence band maximum}

\newacronym{CBM}{CBM}{conduction band minimum}

\newacronym{PBE}{PBE}{Perdew-Burke-Ernzerhof}

\newcommand{\etal}{\textit{et al}.\ }
\newcommand{\ie}{i.e.}

\newacronym{VASP}{\textsc{VASP}}{Vienna \textit{Ab initio} Simulation Package}

\newacronym{SM}{SM}{Supplemental Material}
\newcommand{\sm}{\gls{SM}}
\newacronym{TDA}{TDA}{Tamm-Dancoff approximation}

\newacronym{AHC}{AHC}{Allen, Heine and Cardona}

\newacronym{BZ}{BZ}{Brillouin zone}

\newacronym{MBPT}{MBPT}{Many-body perturbation theory}

\newacronym{CAM}{CAM\ }{Coulomb attenuation method}

\newacronym{GCAM}{GCAM\ }{Generalized Coulomb attenuation method}

\newacronym{MAE}{MAE\ }{Mean absolute error}

\newacronym{HEG}{HEG}{homogeneous electron gas}

\newcommand{\drpa}{\mbox{RPA$\delta_{G,G'}$}}

\begin{document}

\title{Generalized dielectric-dependent hybrid functional combining Gaussian and Yukawa screening}

\author{Alexey Tal}
\email{alexey.tal@vasp.at}
\affiliation{Chaire de Simulation à l'Echelle Atomique (CSEA), Ecole Polytechnique Fédérale de Lausanne (EPFL), CH-1015 Lausanne, Switzerland}
\affiliation{VASP Software GmbH, Berggasse 21, A-1090 Vienna, Austria}
\author{Fabien Tran}
\affiliation{VASP Software GmbH, Berggasse 21, A-1090 Vienna, Austria}
\author{Wei Chen}
\affiliation{Institute of Condensed Matter and Nanoscience (IMCN), Université catholique de Louvain, Louvain-la-Neuve 1348, Belgium}
\author{Stefan Riemelmoser}
\affiliation{Chaire de Simulation à l'Echelle Atomique (CSEA), Ecole Polytechnique Fédérale de Lausanne (EPFL), CH-1015 Lausanne, Switzerland}
\author{Alfredo Pasquarello}
\affiliation{Chaire de Simulation à l'Echelle Atomique (CSEA), Ecole Polytechnique Fédérale de Lausanne (EPFL), CH-1015 Lausanne, Switzerland}

\begin{abstract}
We develop a generalized Coulomb-attenuated method (GCAM) for
dielectric-dependent hybrid functionals that combines Gaussian and Yukawa-type
screening functions to provide an accurate representation of the \rpa dielectric
function. Through systematic fitting to \rpa dielectric functions of
semiconductors and insulators, and to the Lindhard dielectric function of the
homogeneous electron gas, we demonstrate that GCAM significantly reduces errors
in the model dielectric function across the full momentum range. However,
despite this improved physical description of screening, band gap calculations
reveal that GCAM yields larger errors compared to experiment than previous CAM
approaches for wide-gap materials, but smaller errors for most narrow-gap
materials. This counterintuitive result demonstrates that the success of CAM
stems not purely from physical reasons, but also from a partial cancellation of
errors. In contrast, for optical absorption spectra calculated within a model
Bethe-Salpeter equation framework, GCAM systematically outperforms CAM, showing
better agreement with the reference spectra obtained with the diagonal
\gls{RPA} screening in both peak positions and relative intensities. Our findings reveal the important role of error cancellation in
dielectric-dependent hybrid functionals.
\end{abstract}

\maketitle
\section{Introduction}
Density functional theory (DFT) \cite{Hohenberg1964,Kohn1965} has become the
workhorse of computational materials science due to its favorable balance
between accuracy and computational cost. However, the predictive power of DFT
depends critically on the choice of exchange-correlation functional
\cite{Perdew2001,Mardirossian2017,Tran2026}.
Standard local and semilocal functionals, such as the local density approximation (LDA)
and generalized gradient approximation (GGA), systematically underestimate band
gaps of semiconductors and insulators due to the self-interaction error
\cite{Perdew1981}
and the lack of derivative discontinuity in the exchange-correlation potential
\cite{Perdew1983,Sham1983}.

Hybrid functionals \cite{Becke1993,Perdew1996,Adamo1999}, which incorporate a
fraction of Fock exchange, have emerged as a practical solution to this
problem \cite{Bredow2000,Heyd2003,Alkauskas2011}.
 The admixture of Fock exchange reduces self-interaction errors and
significantly improves the description of electronic structure properties,
including band gaps, band structures, and optical properties. Among hybrid
functionals, range-separated hybrids \cite{Bylander1990,Yanai2004,Heyd2003,Yang2023} have attracted
particular attention, as they allow for different treatment of short-range and
long-range exchange interactions. This flexibility is particularly advantageous
for extended systems where long-range and short-range behavior can differ
substantially.

Recent developments of hybrid functionals have improved the accuracy of the
functional considerably, bringing them close to the accuracy of many-body
perturbation theory (MBPT) methods \cite{Hedin1965}. A key insight in this
development is the connection between range-separation schemes and screening
effects. Chen \etal \cite{Chen2018a} have shown that the Coulomb attenuation
used to separate the Coulomb interaction into short- and long-range components
leads to a functional form that is similar to the diagonal form of the
random-phase approximation (RPA). This observation suggests that the range-separation
parameters can be interpreted in terms of physical screening properties of the
material. By fitting the range-separation parameters to the \rpa dielectric
function, the functional was substantially improved, yielding band gaps in
excellent agreement with experiment for a wide range of semiconductors and
insulators.

A similar approach was proposed by Cui \etal \cite{Cui2018}, who used the
same functional form for Fock exchange as in Ref.\ \cite{Chen2018a}. However, the
range-separation parameter was determined from a semi-empirical model dielectric
function proposed by Cappellini \etal \cite{Cappellini1993}. Related developments
include global hybrids with a material-dependent fraction of Fock exchange
\cite{Marques2011,Koller2013}, whose accuracy in non-self-consistent calculations was
assessed in Ref.\ \cite{Skelton2020}, the nonempirical range-separated dielectric-dependent hybrid (DDH)
functionals \cite{Skone2016} and their generalizations
\cite{Wing2017,Zheng2019,Zhan2025} by Galli and coworkers, as well as the
optimally tuned range-separated hybrids \cite{Stein2009,Baer2010,Kronik2012},
which have been extended to solids as screened range-separated hybrids with
the long-range Fock fraction set by the inverse macroscopic dielectric
constant and the range-separation parameter determined by optimal tuning
rather than by fitting to a dielectric function
\cite{RefaelyAbramson2015,Wing2021}.
Another alternative approach was proposed by Riemelmoser \etal \cite{Riemelmoser2026},
where Fock exchange screened via the dielectric constant was combined with
a meta-GGA, yielding an improved description of the screening. A meta-GGA-based
DDH with range separation was also put forward by Jana \etal \cite{Jana2025}.
For a review of hybrid functionals
with system-dependent parameters, see Ref.\ \cite{Zhang2020}. While all these
approaches demonstrate the importance of connecting range-separation to
dielectric screening, they rely on relatively simple model dielectric functions
that may not capture the full complexity of the screening behavior in extended
systems, particularly in the intermediate momentum range.

The success of the DDH functional based on the
Coulomb-attenuating method (CAM) \cite{Chen2018a} raises important
questions about the origin of their accuracy. Is it due to an accurate
description of the underlying physics of screening, or does it rely on
fortuitous error cancellation? Understanding this question is crucial for
further development of DDH functionals and for
extending their applicability to more challenging systems.

In this work, we propose a generalized CAM (GCAM) functional that employs a
flexible model dielectric function, combining both Gaussian and Yukawa-type
screening functions \cite{Yukawa1935,Bylander1990,Akinaga2008}.
We show that this generalization allows for a better representation of the \rpa dielectric
function across the full momentum range. We systematically investigate the
performance of GCAM for band gap predictions and analyze the relationship
between the accuracy of the model dielectric function and the accuracy of
calculated band gaps. Our findings provide new insights into the success of
DDH functionals and point toward future improvements in
functional design.

The remainder of this paper is organized as follows. Section
\ref{sec:theory} presents the theoretical scheme: the GCAM hybrid functional
and the model dielectric function on which it is based
(Sec.\ \ref{sec:gcam}), and the procedure used to determine its parameters
from the \rpa dielectric function (Sec.\ \ref{sec:fitting}). Section
\ref{sec:implementation} describes the practical implementation of the
functional and gives the computational details. Section \ref{sec:results}
presents our main results, beginning with an assessment of the accuracy of
the model dielectric functions for semiconductors and insulators and for the
homogeneous electron gas (HEG) (Sec.\ \ref{sec:accuracy}), followed by band
gap calculations (Sec.\ \ref{sec:bandgaps}) and optical absorption spectra
calculated within a model Bethe-Salpeter equation (BSE) framework
(Sec.\ \ref{sec:spectra}). We conclude in Sec.\ \ref{sec:conclusions}.

\section{Theoretical scheme}
\label{sec:theory}
\subsection{Generalized CAM DDH functional}
\label{sec:gcam}
The two main approaches for Coulomb attenuation used in the literature are based
on error function (erf) \cite{Dombroski1996,Gill1996,Heyd2003,Chen2018a,Cui2018} and Yukawa screening
\cite{Yukawa1935,Bylander1990,Akinaga2008}.
The erf function naturally arises from
Gaussian-type screening in real space, while Yukawa screening corresponds to the
Thomas-Fermi approximation of the dielectric response \cite{Mahan2000}. However,
the idea of Coulomb attenuation can be generalized to a screening
function $S(r)$, where $r=\left\vert\mathbf{r}-\mathbf{r}'\right\vert$
is the interelectronic distance. This function increases monotonically from
$S(0)=0$ to $S(r\to\infty)=1$ and separates the Coulomb interaction into
short-range and long-range components:
\begin{equation}
\frac{1}{r}=\underbrace{\frac{a+b S(r)}{r}}_{\text{Fock}}
+\underbrace{\frac{1-a-b S(r)}{r}}_{\text{semilocal}},
\end{equation}
where the first term is treated with Fock exchange and the second one with
semilocal exchange, and $a$ and $a + b$ denote the fraction of the
short-range and of the long-range parts of Fock exchange, respectively.
Note that the popular Heyd-Scuseria-Ernzerhof (HSE) functional \cite{Heyd2003} corresponds to
$a=-b=0.25$ and $S(r)=\operatorname{erf}\left(\mu r\right)$,
while the screened-exchange LDA (sX-LDA) \cite{Bylander1990} is obtained
with $a=-b=1$ and $S(r)=1-e^{-\mu' r}$, where $\mu$ and $\mu'$ are
suitable range-separation parameters.

Here, we propose a linear combination of erf and Yukawa screening functions:
\begin{equation}
S(r)=\alpha \operatorname{erf}\left(\mu_G r\right)
+(1-\alpha)\left(1-e^{-\mu_Y r}\right),
\label{eq:s}
\end{equation}
where $\alpha$ controls the mixing between the two screening types, and $\mu_G$ and $\mu_Y$ 
are the range-separation parameters for the Gaussian (error function)
and Yukawa (exponential) components, respectively.
As a side remark, we mention that $\operatorname{erf}\left(\mu_G r\right)$
can be rather well approximated by $1-e^{-\mu_Y r}$
if $\mu_Y\approx\left(3/2\right)\mu_G$ \cite{Shimazaki2008,Tran2011,Uzulis2025}.

Setting $a=1$ and $b=\varepsilon^{-1}-1$, where $\varepsilon$ is the macroscopic 
dielectric constant, ensures the correct limiting behavior at long range.
With this choice, the functional contains the full Fock exchange at short
range, where the Fock kernel is $1/r$, and the screened Fock exchange
$1/(\varepsilon r)$ at long range, complemented by the fraction
$1-\varepsilon^{-1}$ of semilocal exchange.
This leads to the following form for the exchange-correlation functional:
\begin{eqnarray}
E_{\mathrm{xc}} & = & E_{\mathrm{x}}^{\mathrm{SR,Fock}} +
(b+1)E_{\mathrm{x}}^{\mathrm{LR,Fock}} - b E_{\mathrm{x}}^{\mathrm{LR,PBE}} +
E_{\mathrm{c}}^{\mathrm{PBE}} \nonumber \\
 & = & E_{\mathrm{x}}^{\mathrm{Fock}} +
b\left(E_{\mathrm{x}}^{\mathrm{LR,Fock}} - E_{\mathrm{x}}^{\mathrm{LR,PBE}}\right)
+ E_{\mathrm{c}}^{\mathrm{PBE}},
\label{eq:exc}
\end{eqnarray}
which uses the Perdew--Burke--Ernzerhof (PBE) GGA functional \cite{Perdew1996a} for
the semilocal part. Given Eq.\ (\ref{eq:s}), the SR and LR parts of the Fock and PBE
exchange functionals read
\begin{equation}
  E_{\rm x}^{\rm SR/LR,Fock} = \alpha E_{\rm x}^{\rm SR/LR\text{-}erf,Fock} + (1-\alpha)E_{\rm x}^{\rm SR/LR\text{-}Yu,Fock}
\label{eq:exlrfock}
\end{equation}
and
\begin{equation}
  E_{\rm x}^{\rm SR/LR,PBE} = \alpha E_{\rm x}^{\rm SR/LR\text{-}erf,PBE} + (1-\alpha)E_{\rm x}^{\rm SR/LR\text{-}Yu,PBE},
\label{eq:exlrpbe1}
\end{equation}
respectively.
Equation \eqref{eq:exc} is the functional used for the calculations
presented in this work. Some details on the screening scheme for the GGA component
are provided in Sec.\ \ref{sec:screened_gga}.

Before addressing the screening in the generalized scheme, we first review
the dielectric function obtained in the original CAM approach. The
DDH functional of Chen \etal \cite{Chen2018a},
denoted DD-RSH-CAM in the original work, adopts the functional form of the
Coulomb-attenuating method \cite{Yanai2004} for the exchange and is referred
to as simply CAM throughout this work. It employs only Gaussian screening
[$\alpha=1$ in Eq.~(\ref{eq:s})], which yields a simple analytical form for the
inverse dielectric function in reciprocal space:
\begin{equation}
\varepsilon^{-1}_{\rm CAM}(q) = 1 + \left( \varepsilon^{-1} - 1 \right) e^{-q^2 / (4\mu^2)},
\label{eq:cam}
\end{equation}
where $\mu$ is the only range-separation parameter and is determined by fitting
to the \rpa dielectric function.

Generalizing this expression to include both Gaussian and Yukawa components 
yields the following GCAM form for the dielectric function:
\begin{equation}
  \varepsilon^{-1}_{\rm GCAM}(q)
  =1+(\varepsilon^{-1}-1)\Big[\alpha e^{-q^2/(4 \mu_G^2)}
  +\frac{(1-\alpha)\mu_Y^2}{q^2+\mu_Y^2}\Big].
\label{eq:gcam}
\end{equation}
The second term in square brackets corresponds to the Thomas-Fermi form of screening,
\begin{equation}
\varepsilon^{-1}_{\rm TF}(q)=1+(\varepsilon^{-1}-1)\frac{\mu_Y^2}{q^2+\mu_Y^2},
\end{equation}
which can be obtained as the small-$q$ ($q \ll 2k_{\rm F}$) limit of the 
Lindhard dielectric function \cite{Lindhard1954}
\begin{equation}
\varepsilon_{\rm L}(q)
=
1 + \frac{k_{\rm TF}^2}{q^2}
\left(
\frac{1}{2}
+
\frac{1 - x^2}{4x}
\ln\left|\frac{1+x}{1-x}\right|
\right),
\end{equation}
where $x = q/(2k_{\rm F})$, $k_{\rm F}=\left(3\pi^2 n\right)^{1/3}$
is the Fermi wave vector, and $k_{\rm TF}=\sqrt{4k_{\rm F}/\pi}$
is the Thomas-Fermi screening wave vector.

However, at finite $q$ values, the Lindhard dielectric function exhibits complex
logarithmic behavior for which a  simple analytical screening function suitable
for range-separation schemes has not been derived, and the inverse dielectric
function (required for the screened exchange) becomes prohibitively complicated.
We therefore represent this finite-$q$ behavior with a Gaussian function,
which has been shown to provide an accurate description of the screening at
intermediate $q$ \cite{Chen2018a,Bokdam2016}.

\begin{figure}[!t]
  \centering
  \includegraphics[width=0.48\textwidth]{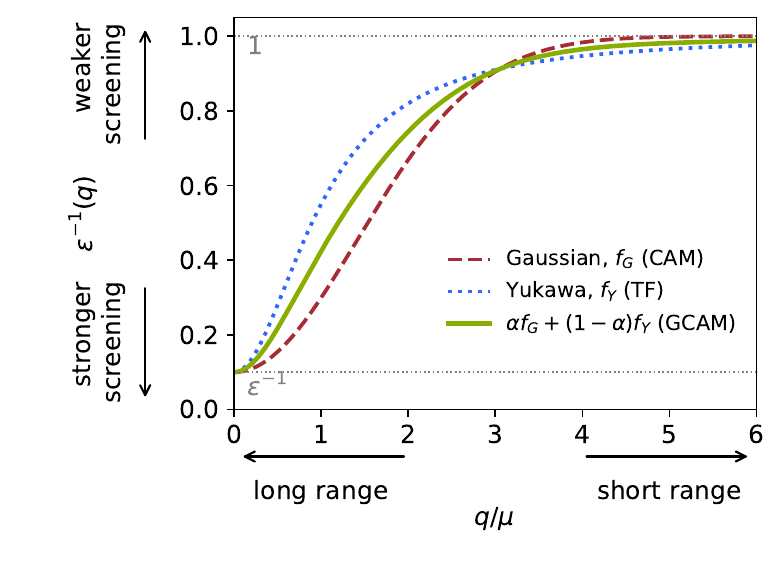}
  \caption{Schematic comparison of the inverse dielectric functions obtained
  with the Gaussian factor $f_G(q)$ alone [CAM, Eq.\ \eqref{eq:cam}], with
  the Yukawa factor $f_Y(q)$ alone (Thomas-Fermi screening), and with their
  mixture [GCAM, Eq.\ \eqref{eq:gcam}], for a common range-separation
  parameter $\mu=\mu_G=\mu_Y$, $\alpha=0.5$, and $\varepsilon=10$. The
  long-range (small $q$) and short-range (large $q$) regions are indicated
  on the horizontal axis. A lower (higher) value of $\varepsilon^{-1}$
  corresponds to stronger (weaker) screening, as indicated on the vertical
  axis.}
  \label{fig:schematic}
\end{figure}

The complementary roles of the two terms in Eq.\ \eqref{eq:gcam} are
illustrated schematically in Fig.\ \ref{fig:schematic}, which compares the
effect of the Gaussian factor $f_G(q)=e^{-q^2/(4\mu^2)}$ of Eq.\ \eqref{eq:cam}
and of the Yukawa factor $f_Y(q)=\mu^2/(q^2+\mu^2)$, together with their mixture
entering Eq.\ \eqref{eq:gcam}, for a common range-separation parameter
$\mu=\mu_G=\mu_Y$. Since each factor multiplies the negative prefactor
$(\varepsilon^{-1}-1)$, a larger factor corresponds to a lower
$\varepsilon^{-1}(q)$, \ie, to stronger screening, and a smaller factor to
weaker screening. At small $q$ (long range), the Yukawa factor decreases as
$f_Y\approx1-q^2/\mu^2$, four times faster than the Gaussian factor
$f_G\approx1-q^2/(4\mu^2)$, so that $\varepsilon^{-1}(q)$ rises steeply from
its $q\to0$ value $\varepsilon^{-1}$, as the \rpa dielectric function does.
With the Gaussian factor alone, as in CAM, $\varepsilon^{-1}(q)$ rises too
slowly, and CAM therefore overestimates the screening at small $q$
(see Fig.\ \ref{fig:fit_cam}). At intermediate $q$, the Yukawa factor alone
would approach zero too slowly, and it is the Gaussian factor that provides
the flexibility needed to describe the crossover of $\varepsilon^{-1}(q)$
toward unity. At large $q$ (short range), $f_G$ vanishes exponentially,
whereas $f_Y$ decays only as $\mu^2/q^2$, so that the asymptotic behavior
of GCAM is set by the Yukawa factor. Neither factor reproduces the $q^{-4}$
decay of the Lindhard function,
$\varepsilon_{\rm L}(q)-1 \approx 4k_{\rm TF}^2k_{\rm F}^2/(3q^4)$: $f_G$
decays too fast, so that $\varepsilon^{-1}_{\rm CAM}(q)$ approaches unity
too quickly and CAM underestimates the screening at large $q$
(Fig.\ \ref{fig:fit_cam}), whereas $f_Y$ decays too slowly.
However, this asymptotic behavior is mostly inconsequential for the hybrid
functional, since at these momenta $\varepsilon^{-1}(q)$ is already close to
unity. Since both limits, $\varepsilon^{-1}$ at $q\to0$ and unity at
$q\to\infty$, are imposed identically in CAM and GCAM, the two models
differ essentially at intermediate $q$.

To maintain the computational simplicity of the original CAM approach
\cite{Chen2018a}, which contains a single fitting parameter, we also introduce a
reduced-parameter version of GCAM (rGCAM). We tie both range-separation lengths to the 
Fermi wave vector through $\mu_G = \mu_Y = \sqrt{\alpha}\,k_{\rm F}$, so that 
$\mu_G^2 = \mu_Y^2 = \alpha\,k_{\rm F}^2$, which yields
\begin{equation}
\begin{split}
\varepsilon^{-1}_{\rm rGCAM}(q)&=1+(\varepsilon^{-1}-1)\bigg[\alpha\, e^{-q^2/(4\alpha k_{\rm F}^2)}\\
&\quad+(1-\alpha)\frac{\alpha k_{\rm F}^2}{\alpha k_{\rm F}^2+q^2}\bigg].
\end{split}
\label{eq:rgcam}
\end{equation}
In this formulation, the mixing parameter $\alpha$ also determines the
range-separation lengths of both terms through their relation to $k_{\rm F}$, 
reducing the model to a single adjustable parameter without significant loss
of accuracy.

\subsection{Determination of hybrid functional parameters}
\label{sec:fitting}
The parameters in the model dielectric functions are determined by fitting to
reference dielectric functions. For semiconductors and insulators, we fit to the
diagonal part of the \rpa dielectric function calculated from PBE eigenvalues and
eigenfunctions. For the HEG, we fit to the analytical
Lindhard dielectric function. 

The macroscopic dielectric constant $\varepsilon$ is obtained from the 
zero-frequency, long-wavelength limit of the \rpa dielectric function for each 
material. For the remaining parameters ($\alpha$, $\mu_G$, $\mu_Y$ for GCAM), 
we minimize the root-mean-square (RMS) deviation
\begin{equation}
{\rm RMS} = \sqrt{\frac{1}{N_q}\sum_{i=1}^{N_q} \left[\varepsilon^{-1}_{\rm
model}(q_i) - \varepsilon^{-1}_{\rm RPA}(q_i)\right]^2}
\end{equation}
between the model and reference inverse dielectric functions over a discrete
set of $N_q$ momentum transfer points.
For semiconductors and insulators, the $q_i$ are the norms
$\left\vert\mathbf{q}+\mathbf{G}\right\vert$ at which the diagonal \rpa
dielectric matrix is computed. Since these points are not uniformly
distributed, we retain only one point per cell of a uniform grid of
spacing 0.08 in the ($q$ in \AA$^{-1}$, $\varepsilon^{-1}$) plane, which
prevents densely populated regions from dominating the fit. For the HEG, the
$q_i$ are 100 uniformly spaced points in the range
$0.01\le q/k_{\rm F}\le 5$.
For the HEG, the fitting is performed separately for each 
density (characterized by the Wigner-Seitz radius
$r_s=\left[3/\left(4\pi n\right)\right]^{1/3}$) to investigate the density-dependent performance 
of the different model forms. Throughout, $r_s$ and
$k_{\rm F}$ are given in atomic units, i.e.\ $r_s$ in units of the 
Bohr radius $a_0$ and $k_{\rm F}$ in units of $a_0^{-1}$.

\section{Practical implementation}
\label{sec:implementation}
\subsection{Treatment of the \texorpdfstring{$q\to0$}{q -> 0} divergence}
\label{sec:singularity}
The Fock exchange energy is evaluated as a Brillouin-zone sum over the
screened kernel $v_{\rm GCAM}(q)=\left(4\pi e^{2}/q^{2}\right)
\varepsilon^{-1}_{\rm GCAM}(q)$. Since
$\varepsilon^{-1}_{\rm GCAM}(0)=\varepsilon^{-1}$, this kernel retains the
full $1/q^{2}$ divergence of the bare Coulomb interaction, merely scaled
by $\varepsilon^{-1}$, and the $q=0$ term of the sum cannot be omitted.
The short-range kernel of HSE \cite{Heyd2003}, by contrast, is regular at
$q=0$ because the screening removes the interaction entirely at long
range. We treat the divergence with the auxiliary-function scheme of
Gygi and Baldereschi \cite{Gygi1986}, in the form implemented in the
Vienna \textit{Ab initio} Simulation Package (VASP) \cite{Kresse1996,Massidda1993,Paier2005}, applied separately to the bare Coulomb,
erf-screened, and Yukawa-screened components of the kernel.

\subsection{One-center Yukawa contribution to the PAW exchange}
The all-electron part of the Fock exchange integrals inside the
projector augmented-wave (PAW)
spheres is evaluated by expanding the screened Coulomb kernel in
spherical harmonics and using an auxiliary radial Green's function
$G_{l}^{\rm SR}(r,r')$. For the Yukawa branch we use the analytic
multipole expansion of the screened Coulomb interaction,
\begin{equation}
  \frac{e^{-\mu_Y|\mathbf r-\mathbf r'|}}
       {|\mathbf r-\mathbf r'|}
  = \mu_Y\sum_{l=0}^{\infty}(2l+1)\,
    i_{l}(\mu_Y r_{<})\,k_{l}(\mu_Y r_{>})\,
    P_{l}(\cos\theta),
\end{equation}
where $i_{l}$ and $k_{l}$ are the modified spherical Bessel functions of
the first and second kinds, respectively, with $r_{<}=\min(r,r')$ and $r_{>}=\max(r,r')$,
and $P_{l}$ are Legendre polynomials.
The radial kernel is therefore
\begin{equation}
  G_{l}^{\rm SR,Yu}(r,r')
  =(2l+1)\,\mu_Y\,
    i_{l}(\mu_Y r_{<})\,k_{l}(\mu_Y r_{>}).
\end{equation}
The
long-range Yukawa part follows from
$G_{l}^{\rm LR,Yu}=r_{<}^{l}/r_{>}^{l+1}-G_{l}^{\rm SR,Yu}$, so that the
bare-Coulomb sum rule is preserved by construction.

The erf branch is treated analogously, with the radial kernel
$G_{l}^{\rm SR,erf}(r,r')$ obtained from the spherical-harmonic expansion
of the error-function-screened Coulomb interaction derived by
\'Ang\'yan \etal~\cite{Angyan2006}. For a mixed
Gaussian+Yukawa kernel the multipole component is assembled as
$\alpha\,G_{l}^{\rm SR,erf}+(1-\alpha)\,G_{l}^{\rm SR,Yu}$, mirroring the
plane-wave kernel.

\subsection{Screened GGA exchange}
\label{sec:screened_gga}
The screening of the semilocal GGA exchange is done as prescribed in
Ref.\ \cite{Heyd2003} for the error function.
It starts from the scaling of the PBE exchange-hole
$J^{\rm PBE}(n,s,y)$ (its expression can be found in
Refs.~\cite{Ernzerhof1998,Heyd2003,Heyd2004}), where 
$s=\left\vert\nabla n\right\vert/\left(2nk_{\rm F}\right)$ is the
reduced-density gradient and $y=k_{\rm F}r$ is the reduced interelectronic distance,
by either $\operatorname{erfc}(\eta_G y)$ or $e^{-\eta_Y y}$
(where $\eta_{G/Y}=\mu_{G/Y}/k_{\mathrm F}$), leading to the SR PBE enhancement factors
\begin{equation}
\begin{split}
  F_{\rm x}^{\rm SR\text{-}erf,PBE}&(n,s,\mu_G)\\
  &=-\tfrac{8}{9}\int_{0}^{\infty}\!y\,J^{\rm PBE}(n,s,y)\,\operatorname{erfc}(\eta_G y)\,{\rm d}y
\end{split}
\label{eq:fxerf}
\end{equation}
and
\begin{equation}
\begin{split}
  F_{\rm x}^{\rm SR\text{-}Yu,PBE}&(n,s,\mu_Y)\\
  &=-\tfrac{8}{9}\int_{0}^{\infty}\!y\,J^{\rm PBE}(n,s,y)\,e^{-\eta_Y y}\,{\rm d}y
\end{split}
\label{eq:fxyu}
\end{equation}
for the erf and Yukawa screenings, respectively.
The integration in Eq.\ (\ref{eq:fxerf}) is done semi-analytically \cite{Heyd2003,Heyd2004},
while in the case of Eq.\ (\ref{eq:fxyu}) the integration is performed fully numerically
due to the absence of a closed-form expression.

Then, the SR PBE exchange energy is calculated with
\begin{equation}
\begin{split}
  E_{\rm x}^{\rm SR\text{-}erf/Yu,PBE} &= \int
  \epsilon_{\rm x}^{\rm LDA}(n)\\
  &\quad\times F_{\rm x}^{\rm SR\text{-}erf/Yu,PBE}(n,s,\mu_{G/Y})\,{\rm d}^3r,
\end{split}
\label{eq:exsrpbe}
\end{equation}
where $\epsilon_{\rm x}^{\rm LDA}(n)=-\left(3/4\right)\left(3/\pi\right)^{1/3}n^{4/3}$
is the exact expression for the HEG.

The corresponding LR parts, used for evaluating Eqs.\ (\ref{eq:exc}) and (\ref{eq:exlrpbe1}),
are calculated as
\begin{equation}
  E_{\rm x}^{\rm LR\text{-}erf/Yu,PBE} = E_{\rm x}^{\rm PBE} - E_{\rm x}^{\rm SR\text{-}erf/Yu,PBE}.
\label{eq:exlrpbe2}
\end{equation}
Note that the implementation of Eq.\ (\ref{eq:exlrpbe2}) in VASP
uses the original form of the PBE exchange functional \cite{Perdew1996}
for the $E_{\rm x}^{\rm PBE}$ term, while other implementations may use the
formulation based on the hole $J^{\rm PBE}$ \cite{Heyd2004}.
The resulting screened DDH functional reduces to plain PBE in the
$\mu_{\mathrm G}\to0$ and $\mu_{\mathrm Y}\to0$ limit.

\subsection{Computational details}
All DFT calculations were performed using VASP \cite{Kresse1996}.
We employed the PAW method \cite{Blochl1994,Kresse1999} to describe the 
electron-ion interaction, with PAW potentials from the VASP database. 

Band gap calculations using the DDH functionals (CAM 
and GCAM) were performed self-consistently, with the screening 
parameters determined from the \rpa dielectric functions as described in 
Sec.\ \ref{sec:fitting}. Convergence criteria were set to $10^{-6}$ eV for 
electronic self-consistency.
The plane-wave kinetic energy cutoff and $\mathbf{k}$-point mesh are given in
Table~S1 of the \sm\ \cite{SM}.

Initial ground-state calculations were performed using the PBE exchange-correlation
functional \cite{Perdew1996a} to obtain converged charge densities and
Kohn-Sham eigenvalues. Experimental lattice parameters were used for all 
materials to facilitate comparison with experimental band gaps,
see Table~S2 in the \sm\ \cite{SM}.

\section{Results}
\label{sec:results}
\subsection{Accuracy of model dielectric functions}
\label{sec:accuracy}

\subsubsection{Semiconductors and insulators}
\begin{figure}[!hbpt]
  \centering
  \includegraphics[width=0.48\textwidth]{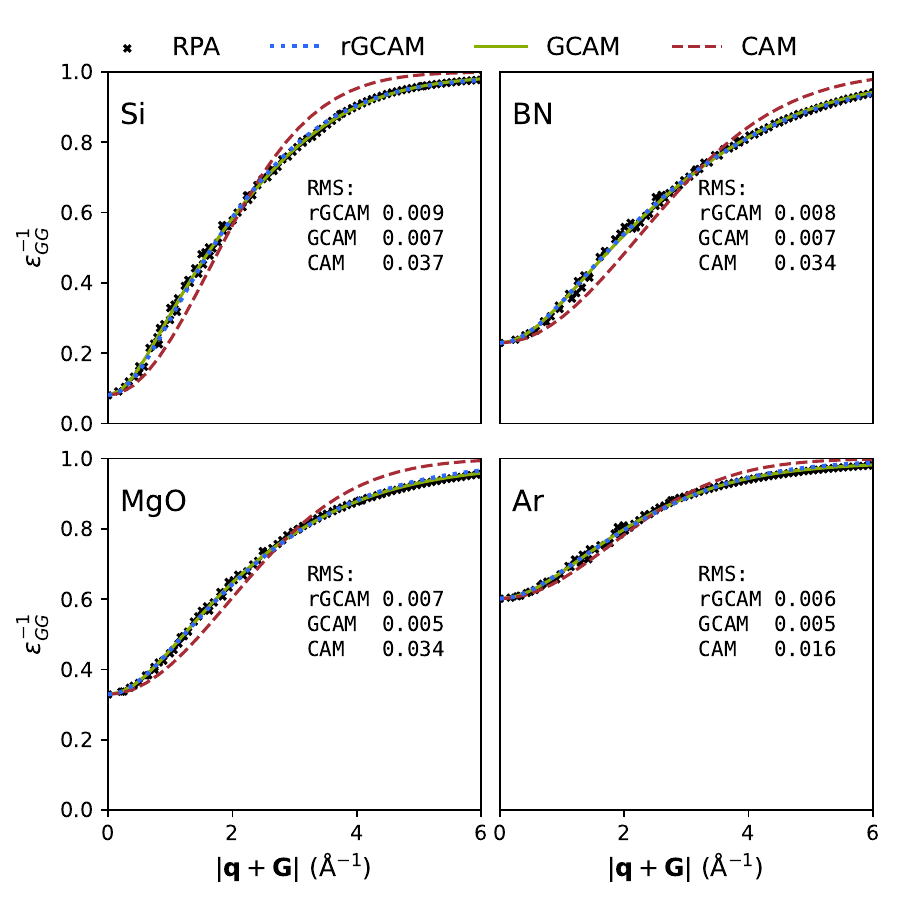}
  \caption{Fitting of model dielectric functions (CAM, GCAM, rGCAM) to \rpa 
  dielectric functions for representative semiconductors and insulators. The 
  comparison shows systematic improvement of GCAM over CAM across all materials,
  with RMS errors reduced by 70--85\%.}
  \label{fig:fit_cam}
\end{figure}

Our benchmark set consists of the 14 semiconductors and insulators listed in
Table \ref{tab:bandgaps}, ranging from Si to LiF. Figure \ref{fig:fit_cam}
shows the results of fitting CAM, GCAM, and rGCAM to the \rpa dielectric
function for four of these materials (Si, BN, MgO, and Ar), chosen as a
representative sample. The fits for the full set are given in Fig.~S1 of the
\sm\ \cite{SM}, and the resulting parameters are listed in Table~S3.
The RMS errors demonstrate that GCAM provides significantly
improved accuracy over CAM across all materials studied.
With a single parameter, rGCAM yields errors that are in most cases only
slightly larger than those of GCAM and consistently smaller than those of CAM.
rGCAM thus offers a simplified alternative to GCAM and, judging from the
accuracy of the fits, is expected to yield very similar results. In the
following, we focus on the more general GCAM.

Detailed analysis of the fitted curves reveals the origin of these improvements.
CAM systematically overestimates the screening at small $q$ (long range) while
underestimating it at intermediate to large $q$ (short range). This asymmetric
error is a consequence of using only a Gaussian function to represent the full
momentum dependence of the dielectric response. In contrast, GCAM captures the
correct behavior across the full $q$ range through the combination of Gaussian
and Yukawa components, with the Yukawa term providing the proper long-range
Thomas-Fermi-like screening and the Gaussian term accurately representing the
crossover toward unity at intermediate $q$ (see Sec.\ \ref{sec:gcam}).

\subsubsection{Homogeneous electron gas}
\begin{figure}[!hbpt]
  \centering
  \includegraphics[width=0.48\textwidth]{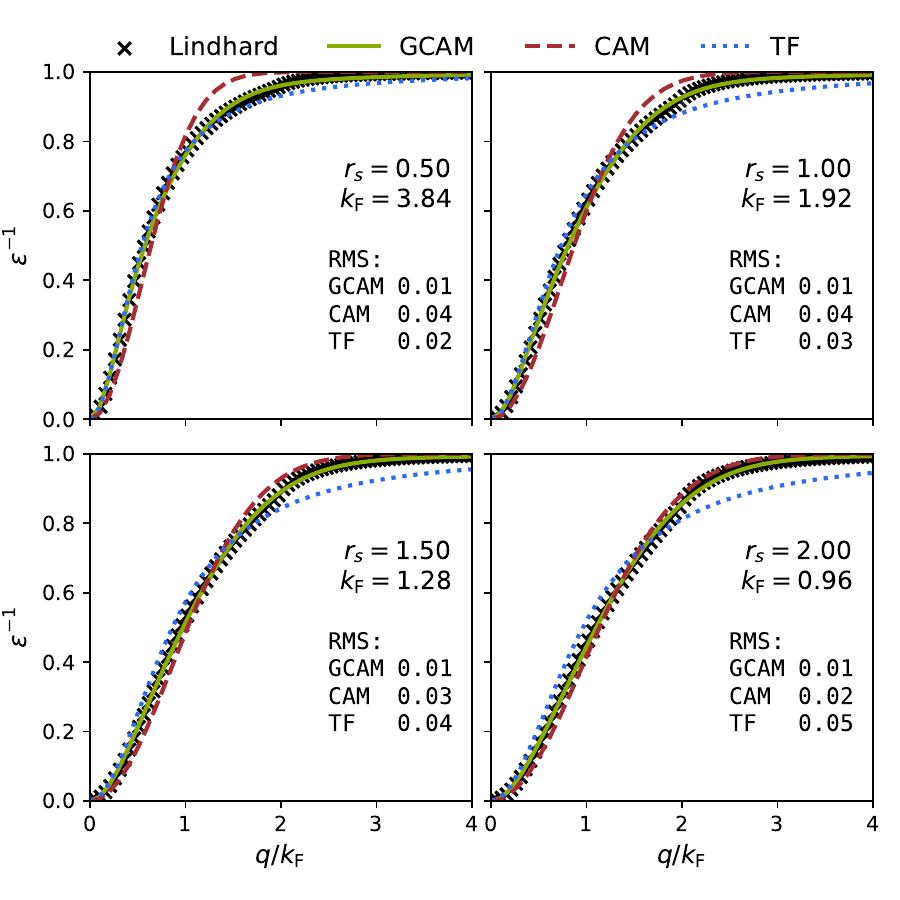}
  \caption{Fitting of model dielectric functions to the Lindhard dielectric
  function of HEG for densities ranging from
  $r_s = 0.5$ to $r_s = 2.0$. The Thomas-Fermi (TF) model, \ie, pure Yukawa
  screening, is shown for comparison. The comparison demonstrates the necessity of
  both Gaussian and Yukawa components for accurate description across different
  density regimes.}
  \label{fig:fit_heg}
\end{figure}

To understand the performance of the model dielectric functions across different
charge density regimes, we fit them to the analytical Lindhard dielectric
function for HEG at various densities $r_s$.
Figure \ref{fig:fit_heg} shows results for $r_s$ values ranging from 0.5 (high
density) to 2.0 (low density), spanning the typical range relevant for real
materials. Results for other values of $r_s$ are shown in Fig.~S2 of the SM \cite{SM}.

The comparison reveals a clear transition between two regimes. At high densities
(small $r_s$), the Yukawa component dominates the screening behavior, and models
based solely on the error function (Gaussian) systematically underestimate the
screening in the short range. Conversely, at low densities (large $r_s$), the
error function provides a superior description, while the Yukawa function alone
fails to capture the correct screening behavior. This density-dependent crossover
demonstrates that both functional forms are essential for a universally applicable
model dielectric function.

It is worth noting that the analytical model dielectric functions (CAM and GCAM)
do not capture the non-analytic behavior of the Lindhard function at $q = 2k_{\rm F}$,
which manifests as a kink in the exact response. This singularity arises from the
sharp Fermi surface of the electron gas and cannot be reproduced by smooth
analytical models. However, it is specific to metals: in semiconductors and
insulators the band gap removes the Fermi surface, and the \rpa dielectric
function is smooth at all $q$ (see Fig.\ \ref{fig:fit_cam}). The absence of this
feature in the model functions is therefore inconsequential for the systems
considered here.

\subsection{Band gaps}
\label{sec:bandgaps}
Table \ref{tab:bandgaps} presents the calculated band gaps for a representative 
set of semiconductors and insulators, comparing CAM and GCAM results with 
experimental values corrected for zero-point renormalization (ZPR). The comparison 
reveals a systematic trend: GCAM consistently yields larger band gaps than CAM
across the entire set, shifting every gap upward by 0.12--0.35 eV.

For the narrow- and intermediate-gap semiconductors (Si through BN), CAM tends to
slightly underestimate the gaps, the exception being ZnS, whereas the upward
shift introduced by GCAM brings several of these materials closer to experiment
(e.g., for Si the error changes from $-0.16$ eV with CAM to $-0.02$ eV with GCAM, and
for BN from $-0.19$ eV to $+0.04$ eV).
For the wide-gap insulators (CaO, MgO,
Ar, LiF), however, both functionals overestimate the gaps, and the
additional upward shift causes GCAM to overshoot substantially more than CAM. The
error for LiF grows from $+0.90$ eV (CAM) to $+1.16$ eV (GCAM) and for MgO from
$+0.23$ eV to $+0.58$ eV. These wide-gap overestimations dominate the overall error
of GCAM.

The overall performance metrics quantify this behavior. CAM achieves a mean
absolute error (MAE) of 0.24 eV with a small mean signed error (MSE) of
0.08 eV: because CAM underestimates the narrow-gap semiconductors and overestimates
the wide-gap insulators, the errors are of both signs and largely cancel in the mean.
In contrast, GCAM shows a larger MAE of 0.30 eV with an MSE of 0.29 eV---essentially
all errors are positive, reflecting systematic overestimation with no cancellation.
In relative terms, the picture is different: the mean absolute
percentage error (MAPE) is 4.5\% for CAM and 3.9\% for GCAM, since the
underestimation of the narrow-gap materials by CAM weighs more heavily on a
percentage scale.

This counterintuitive result---that a more accurate model dielectric function 
yields less accurate band gaps---can be related to the fitted dielectric
functions in Sec.\ \ref{sec:accuracy}. Since CAM and GCAM share the same
limits, $\varepsilon^{-1}$ at $q\to0$ and unity at $q\to\infty$, the two
functionals differ only in how the screening interpolates between them, i.e.,
at intermediate $q$. As shown in Fig.\ \ref{fig:fit_cam}, in this range CAM
overestimates the screening at the lower end and underestimates it at the
upper end, whereas GCAM closely follows the \rpa dielectric function.
Correcting the intermediate-range screening shifts all gaps upward. Since the gaps
increase for all materials, the reduced screening at the lower end of this
range outweighs the enhanced screening at the upper end, consistent with the
$1/q^2$ weighting of the Coulomb kernel, which emphasizes smaller momenta. For the narrow-gap semiconductors, which CAM 
underestimates, this shift is partly beneficial, but for the wide-gap insulators it 
amplifies an already-present overestimation. The near-zero MSE of CAM therefore 
reflects a fortuitous cancellation between underestimated narrow gaps and 
overestimated wide gaps rather than a uniformly accurate description.

The success of CAM for band gap predictions thus stems not purely from physical
reasons, but also from a partial cancellation of errors. This finding has important implications
for functional development: improving one aspect of the underlying physics does not
necessarily improve all predicted properties.

The systematic overestimation of the band gaps found with GCAM also indicates
that effects not contemplated in the present DDH approach act to reduce the
gaps, in particular for the wide-gap insulators. In the DDH scheme,
screened Fock exchange in Eq.\ \eqref{eq:exc} is complemented by
semilocal PBE exchange, Eq.\ \eqref{eq:exlrpbe1}, which is screened in the
same way and is meant to account for the part of the exchange-correlation
energy that is not described by screened exchange. Once the screening is
modified at intermediate $q$, however, this semilocal counterpart no longer
accounts for the correct local term. In the COHSEX approximation of
many-body perturbation theory \cite{Hedin1965}, screened exchange is
accompanied by a static and local Coulomb-hole (COH) term, which depends
explicitly on the screened interaction. A more systematic approach would
therefore be to derive the local counterpart of the GCAM screened exchange
as a COH term, following the scheme of Shimazaki and Asai
\cite{Shimazaki2008,Shimazaki2009}, in which such a term is obtained from a
model dielectric function. In addition, the model dielectric functions are
strictly diagonal, so that the off-diagonal elements of the dielectric
matrix, \ie, local-field effects in the screening, are completely neglected
in GCAM. The comparison between diagonal and full \rpa screening in
Sec.\ \ref{sec:spectra} shows that these terms can affect the exciton
binding energies by up to 0.19~eV, but their effect on the band gaps remains
to be assessed. In turn, the present results thus also provide
clear directions for further improvement, which we leave for future work.

\begin{table}[!h]
\caption{\label{tab:bandgaps}Calculated band gaps (in eV) for semiconductors 
and insulators using CAM and GCAM model dielectric functions, compared with 
experimental values corrected for zero-point renormalization (ZPR).
Experimental values are taken from the compilation of
Ref.~\cite{Chen2018a} and from
Refs.~\cite{Bludau1974,Agui1999,Clark1964,Chrenko1974,%
Humphreys1981,Whited1973,BaldiniBosacchi1970,Piacentini1976,RoesslerWalker1968,%
Baldini1962} and references therein. The ZPR corrections are taken from
Refs.~\cite{Miglio2020,Shang2021,Moerman2025,Riemelmoser2026}
and are listed together with their sources for each material in Table~S4 of
the \sm\ \cite{SM}. The table
includes mean absolute error (MAE), mean signed error (MSE), and mean absolute
percentage error (MAPE) relative to experiment.}
\centering
\begin{tabular*}{\columnwidth}{@{\extracolsep{\fill}}lccc}
\toprule
   & CAM & GCAM & Expt+ZPR\\
\midrule
Si & 1.07 & 1.21 & 1.23\\
BP & 2.00 & 2.16 & 2.21\\
GaP & 2.34 & 2.52 & 2.42\\
SiC & 2.46 & 2.65 & 2.60\\
AlP & 2.47 & 2.63 & 2.61\\
ZnS & 4.08 & 4.35 & 3.93\\
C & 5.68 & 5.90 & 5.81\\
BN & 6.62 & 6.85 & 6.81\\
CaO\footnote{Direct gap at $\Gamma$. See discussion in Ref.\ \cite{Franckel2025}.} & 7.81 & 8.10 & 7.51\\
MgO & 8.58 & 8.93 & 8.35\\
NaCl & 9.24 & 9.42 & 9.27\\
LiCl & 9.85 & 10.06 & 9.84\\
Ar & 14.96 & 15.08 & 14.32\\
LiF & 16.23 & 16.49 & 15.33\\
\addlinespace
MAE & 0.24 & 0.30 & \\
MSE & 0.08 & 0.29 & \\
MAPE & 4.5 & 3.9 & \\
\bottomrule
\end{tabular*}
\end{table}

\subsection{Optical absorption spectra}
\label{sec:spectra}
While GCAM does not uniformly improve the band gap predictions compared to CAM,
the improved accuracy of the model dielectric function has important implications
for optical properties. The calculation of optical absorption spectra within the
Bethe-Salpeter equation (BSE) formalism \cite{Salpeter1951,Onida2002,Rohlfing1998} requires an
accurate description of the screened Coulomb interaction, which enters the
electron-hole kernel. In a model BSE (mBSE) approach, the computationally
expensive \emph{ab initio} screening from $GW$ calculations is replaced by a model
dielectric function, significantly reducing the computational cost while
maintaining high accuracy \cite{Bokdam2016,Tal2020}. 

\begin{figure}[!tb]
  \centering
  \includegraphics[width=0.50\textwidth]{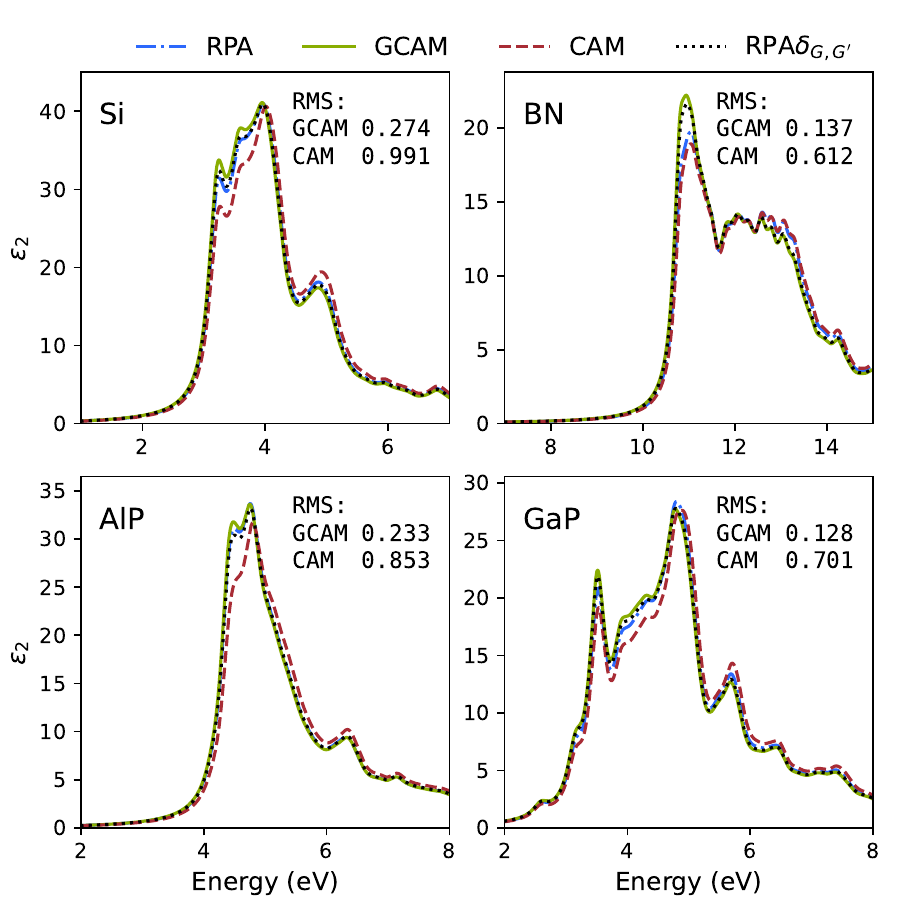}
  \caption{Optical absorption spectra calculated using different model dielectric
  functions for the screening in BSE calculations. Comparison between CAM (red,
  dashed), GCAM (green, solid), the diagonal \rpa reference \drpa{}
  (black, dotted), and the full \rpa screening (blue, dash-dotted) shows that GCAM
  provides better agreement with the diagonal \rpa reference for both peak
  positions and relative intensities.}
  \label{fig:spectra}
\end{figure}

Figure \ref{fig:spectra} shows the optical absorption spectra calculated using
different model dielectric functions for the screening in the BSE calculations.
The spectra for all systems are shown in Fig.~S3 of the SM \cite{SM}.
All spectra are computed starting from $G_0W_0$@PBE quasiparticle energies and 
PBE wave functions.
We employ the Tamm-Dancoff approximation \cite{Tamm1945,Dancoff1950,Onida2002},
which neglects the coupling between resonant and antiresonant transitions, to
solve the BSE. The comparison between CAM, GCAM, and full \rpa screening reveals
several important features. 

Since the model dielectric functions are strictly diagonal and do not account
for the off-diagonal contributions to the screening, we use the diagonal part of
the \rpa dielectric matrix, denoted \drpa{}, as a reference to evaluate their
performance. A comparison with the full \rpa screening would mix the error of
the model dielectric function with the effect of the neglected off-diagonal
elements, which neither model can capture, whereas the comparison with
\drpa{} isolates the quality of the model itself and excludes any
fortuitous compensation between the two. GCAM provides a systematically better approximation to
\drpa{} than CAM across the entire frequency range. This improvement
is reflected in the better agreement of peak positions and relative intensities
with the reference BSE results.

In particular, the figure demonstrates that GCAM captures the fine structure of
the absorption spectra more accurately than CAM. The improved description of
the screening at intermediate $q$ allows GCAM to better
reproduce the excitonic effects that dominate the low-energy part of the
spectrum. For higher-energy transitions, GCAM also shows improved agreement with
\drpa{}, suggesting that the more flexible functional form provides a
better representation of the momentum-dependent screening.

The comparison between \drpa{} and the full \rpa screening, which
additionally includes the off-diagonal elements of the dielectric matrix,
\ie, local-field effects in the screening, allows us to assess the limitations inherent to
any diagonal model of the screening, irrespective of its functional form.
For systems such as Si, AlP, and BP, where the screening is diagonally
dominant, the diagonal description yields excellent results. For systems
like LiF, BN, or C, by contrast, small discrepancies between the diagonal and the
full \rpa screening become visible. For Si, AlP, and GaP, GCAM is much
closer than CAM to both the \drpa{} and the full \rpa spectra. Where the
diagonal approximation itself deviates from the full \rpa result, as for BN,
the weaker excitonic enhancement resulting from the overestimated screening
of CAM can bring its spectrum closer to full \rpa than \drpa{} or GCAM. This
agreement is fortuitous: it does not reflect a better description of the
off-diagonal screening, which neither model contains. The effect of the off-diagonal elements is also apparent in
the exciton binding energies given in Table \ref{tab:exciton}, where the
diagonal description deviates from the full \rpa result by up to 0.19~eV
(LiF), in either direction: the binding energy is overestimated for LiF and
MgO but underestimated for Ar, LiCl, NaCl, and CaO. Nevertheless, for most
semiconductors the diagonal description remains very accurate, and the
improved GCAM model overall shows better agreement with \drpa{} than CAM.

\begin{table}[!h]
\caption{\label{tab:exciton}Calculated exciton binding energy (in eV) for
semiconductors and insulators using BSE with screening described by \gls{RPA},
\drpa{}, and model dielectric functions CAM and GCAM. The table
includes mean absolute error (MAE) relative to \drpa{}.}
\centering
\begin{tabular*}{\columnwidth}{@{\extracolsep{\fill}}lcccc}
\toprule
     & \rpa & \drpa & CAM & GCAM\\
\midrule
Si & 0.084 & 0.084 & 0.084 & 0.085\\
GaP & 0.104 & 0.104 & 0.104 & 0.105\\
AlP & 0.136 & 0.136 & 0.136 & 0.137\\
BP & 0.143 & 0.144 & 0.143 & 0.145\\
ZnS & 0.185 & 0.186 & 0.185 & 0.187\\
SiC & 0.203 & 0.202 & 0.201 & 0.203\\
C & 0.288 & 0.289 & 0.289 & 0.291\\
CaO & 0.432 & 0.419 & 0.407 & 0.421\\
BN & 0.431 & 0.429 & 0.419 & 0.434\\
MgO & 0.497 & 0.513 & 0.506 & 0.515\\
LiCl & 0.793 & 0.761 & 0.722 & 0.773\\
NaCl & 0.838 & 0.821 & 0.796 & 0.826\\
Ar & 1.963 & 1.909 & 1.890 & 1.921\\
LiF & 1.941 & 2.117 & 2.059 & 2.135\\
MAE &    &    & 0.012 & 0.005\\
\bottomrule
\end{tabular*}
\end{table}

\section{Conclusions}
\label{sec:conclusions}
We have developed a generalized Coulomb-attenuated method (GCAM) for
DDH functionals that combines Gaussian and Yukawa-type
screening functions. Systematic comparison with the \rpa dielectric function
demonstrates that GCAM provides significantly improved accuracy over the standard
CAM approach across the full momentum range: the Yukawa scheme reproduces the
Thomas-Fermi limit of the screening at long range (small $q$), while the
Gaussian scheme describes the crossover of the screening toward unity at
intermediate $q$.

Despite this improved physical description of screening, our band gap
calculations reveal a counterintuitive result: GCAM yields larger errors
compared to experiment than CAM for wide-gap materials, but smaller errors for
most narrow-gap materials. This demonstrates that CAM's accuracy stems not purely from a physically
accurate representation of the dielectric function, but also from a partial
cancellation of errors between underestimated short-range and overestimated
long-range screening. Since both functionals share the
same long- and short-range limits, the difference between them originates
from the screening at intermediate range.

For optical absorption spectra calculated within a model BSE
framework, GCAM systematically outperforms CAM in reproducing the spectra
obtained with the diagonal \gls{RPA} screening, with improved agreement in both
peak positions and relative intensities. A comparison with the full \rpa
screening, which includes the off-diagonal elements, reveals that the error
cancellation of CAM can also occur for certain optical spectra. Overall, we
conclude that the error cancellation of CAM depends significantly on the
specific material.

\bibliography{article}
\end{document}